\documentclass[twocolumn,preprintnumbers,amsmath,amssymb]{revtex4}

\usepackage{graphicx}
\usepackage{subfigure}
\usepackage{dcolumn}
\usepackage{bm}
\begin{document}
\title{First-principles modeling of BaCeO$_{3}$:  structure and 
stabilization of O vacancies by Pd-doping}
\author{Joseph W. Bennett$^{a}$, Ram Seshadri$^{b}$, Susannah L.
Scott$^{c}$ and Andrew M. Rappe$^{a}$.}  \affiliation{ (a)The Makineni
Theoretical Laboratories, Department of Chemistry,\\ University of
Pennsylvania, Philadelphia, PA 19104--6323.\\ (b) Materials Department
and Materials Research Laboratory,\\ University of California, Santa
Barbara, CA 93106\\ (c) Department of Chemical Engineering and
Department of Chemistry and Biochemistry, \\University of California,
Santa Barbara, CA 93106} \date{\today}

\begin{abstract}

We use first-principles density functional theory (DFT) calculations
to investigate the ground state structures of both BaCeO$_{3}$ (BC)
and Pd-doped BC (BCP).  The relaxed structures match closely with
recent experimental scattering studies, and also provide a local
picture of how the BC perovskite lattice accommodates Pd.  Both
stoichiometric and oxygen-deficient materials are considered, and
structures with an O vacancy adjacent to each Pd are predicted to be
favored.  The oxidation state of Pd in each doped structure is
investigated through a structural analysis, the results of which are
supported by an orbital-resolved projected density of states.  The
vacancy stabilization by Pd in BCP is explained through redox
chemistry and lattice strain relief.

\end{abstract}

\maketitle
\section{\label{sec:level1} Introduction}
The structure and composition of solid oxide perovskites have been varied
widely, in order to tailor their properties for
applications~\cite{Ho93p4497,Iniguez01p095503,Akbas01p123,Pena01p1981,Ikawa03p2511,Grinberg04p1760,Gomez05p094703,Suchomel05p262905,Haumont05p104106}.
The development of perovskites as either catalysts themselves or as better
oxide supports for catalytic applications is the focus of great current
interest.~\cite{Choi07p25,Erri06p5328,Moreau06p362}

Nishihata {\em et al.} reported~\cite{Nishihata02p164,Tanaka06p5998} that when
precious metals (Pt, Rh, or Pd) are mixed into solid oxides of La(Fe,Co)O$_3$,
then sintered, the precious metal becomes incorporated within the host
perovskite as an oxidized species. When subjected to a reducing atmosphere,
the precious metal is extruded from the host matrix and deposited on the
surface of the oxide as a fully reduced metal. According to Nishihata {\em et
al.} this metal is the active species for catalytic reactions, such as the
oxidation of CO to CO$_{2}$. When the reaction is complete, the catalyst is
subject to an oxidizing atmosphere. This oxidizes the precious metal, which
returns back into the host oxide, where it can be stored for future catalysis.
The two-stage cycle can be repeated many times.  This increases the efficiency
of their catalyst system, presumably because incorporation inhibits precious
metal agglomeration under oxidizing conditions.  As a result, the catalytic
metal retains higher surface area for much longer than on conventional
catalytic substrates.  This work in so-called ``intelligent catalysts'' has
focused mainly on La-based perovskite hosts, where precious metal dopant
levels of up to 5\% were achieved.

Based on crystal chemical size rules, we investigated whether the
upper limit of 5\% precious metal doping is related to the size
mismatch between Pd and Fe.  For analysis of ionic size in
perovskites, the tolerance factor $t$ is the central parameter:

\begin{eqnarray}
 t = \frac {R_{\rm A-O}}  {R_{\rm B-O} \sqrt 2}
\end{eqnarray}            

\noindent where $R_{\rm A-O}$ is the sum of A and O ionic radii and $R_{\rm
B-O}$ is the sum of B and O ionic radii.  LaFeO$_3$ ($r_{\rm La}$=1.36~\AA,
$r_{\rm Fe}\approx$0.6~\AA, $r_{\rm O}$=1.42~\AA) has a tolerance factor of
0.97.  The introduction of a Pd$^{+2}$ on the B site ($r_{\rm Pd}$=0.86~\AA)
would locally make $t=0.86$ which would create significant strain and limit
solubility.  Pd$^{+4}$ is a smaller ion but perhaps a more difficult oxidation
state to stabilize.

To improve on the La-based materials and to find new oxide functionality, we
explored the ability of other materials to incorporate precious metals
reversibly. One important modification scheme has been the use of large A-
and B-site cations, to increase the unit cell volume.  It has been shown that
large cations can yield perovskite oxides that function well as either proton
or oxygen ion conductors~\cite{Kreuer04p4637,Shimura05p2945}, the most
promising being doped BC. The presence of large ions ($r_{\rm Ba}$=1.61~\AA\
and $r_{\rm Ce}$=0.87~\AA) on both A- and B-sites makes the host perovskite
cell unusually large, and we propose that this large volume can more readily
accommodate precious metal ions.

Another important perovskite design strategy pursued in this work is the use
of an ionic, low-valence A-site cation (Ba), in order to stabilize higher
redox states in adjacent cations, in this case the cation incorporation of Pd.
The large size of BC and the redox-stabilization of Ba make BC a suitable host
for precious metal dopants.  We recently
investigated~\cite{Li07p1418,Singh07p347} how these design principles apply to
Pd-doped BC, namely whether BCP reversibly incorporates Pd and whether it is
an active catalyst.

BaCeO$_{3}$ has $t$=0.94 and is predicted to be an
anti-ferrodistortive dielectric material.  Since Ce$^{+4}$ and
Pd$^{+2}$ have the same ionic radii in 6-fold coordination, Pd doping
maintains $t$ and should not induce much strain.  To maintain charge
neutrality, Pd$^{+2}$ would need to be accompanied by an oxygen
vacancy.

In this study, we examine the atomic and electronic properties of Pd-doped BC
in detail using DFT.  The local bonding environments around Pd as well as the
electron fillings of atomic orbitals on Pd and its neighbors are examined.  A
particular focus is the character of the highest occupied molecular orbital
(HOMO) and the few lowest unoccupied molecular orbitals (LUMO).  These
electronic states are compared with expectations based on crystal field
splittings in the computed atomic geometries.  This combined structural and
electronic computational study helps explain the Pd incorporation previously
reported.

There are several theoretical studies of pure BaCeO$_{3}$ and with various
dopants~\cite{Glockner99p145,Katahira00p91,Davies99p323,Wu05p846}, but none of
these reported theoretical predictions of the full ground state structure and
symmetry group of pure BaCeO$_{3}$ using DFT.  Our preliminary theoretical
studies are the first to consider Pd doping of BaCeO$_{3}$, which agree well
with recently published experimental data~\cite{Li07p1418,Singh07p347}. In
this paper we investigate the atomic and electronic structure of BaCeO$_{3}$
and BaCe$_{1-x}$Pd$_{x}$O$_{3}$ in more depth, building upon previously
reported results.

\section{\label{sec:level1} Methodology}
In this study, an in-house solid state DFT code employed in previous
studies~\cite{Mason04p161401R,Bennett06p180102R} is used to relax the ionic
positions and optimize lattice constants. The generalized gradient
approximation (GGA)~\cite{Perdew96p3865} of the exchange correlation
functional and a $4\times4\times4$ Monkhorst-Pack sampling of the Brillouin
zone~\cite{Monkhorst76p5188} are used for all calculations. All atoms are
represented by norm-conserving optimized~\cite{Rappe90p1227} designed
nonlocal~\cite{Ramer99p12471} pseudopotentials. All pseudopotentials are
generated by the OPIUM code~\cite{Opium}. The calculations are performed with
a plane wave cutoff of 50~Ry. Undoped and Pd-doped BaCeO$_{3}$ calculations
were performed on forty atom unit cells in a $2\times2\times2$ supercell
arrangement allowing for the three-dimensional octahedral tilt that is seen in
experiment. In the Pd-doped perovskite, two supercells of composition
BaCe$_{0.875}$Pd$_{0.125}$O$_{2.875}$ were studied, as well as the structure
containing no vacancy, BaCe$_{0.875}$Pd$_{0.125}$O$_{3}$. Structure 1 contains
an O vacancy adjacent to Pd, and Structure 2 contains the vacancy in the next
unit over, between two Ce. Structure 3 contains no vacancy.

\section{\label{sec:level1} Results}
In the forty-atom BaCeO$_{3}$ $2\times2\times2$ unit cell, the O$_{6}$ tilt
angles are all 12.2$^{\circ}$. However, the tilts along (100) and (001) are
anti-phase and indistinguishable from each other, while the tilts along (010)
are in-phase. This structure belongs to the $a^{-}a^{+}a^{-}$ tilt system. The
converged structure belongs to the $Pnma$ symmetry group, and the forty-atom
computational cell we used is a supercell of the primitive
$\sqrt2\times2\times\sqrt2$ cell. Relaxed DFT lattice constants agree well
with experiment~\cite{Li07p1418}, as seen in Table~\ref{table:BaCeO3Param}, as
do the positions of ions from the experimental refinements.

\begin{table}
\begin{tabular}{c|c|c}
\hline
&Experiment&Theory\\
\hline
$a$&6.21&6.28\\
$b$&8.77&8.81\\
$c$&6.23&6.30\\
\hline
Ba($x$)&0.0207&0.0273\\
Ba($z$)&-0.0075&-0.0077\\
O1($x$)&-0.0132&-0.020\\
O1($z$)&0.4250&0.4194\\
O2($x$)&0.2771&0.2833\\
O2($y$)&0.0383&0.0446\\
O2($z$)&0.7239&0.7184\\
\hline
\end{tabular}
\caption{Comparison of experimental~\cite{Li07p1418} and theoretical structural
parameters for BaCeO$_{3}$.}
\label{table:BaCeO3Param}
\end{table}

\begin{figure}
\includegraphics[width=3.0in]{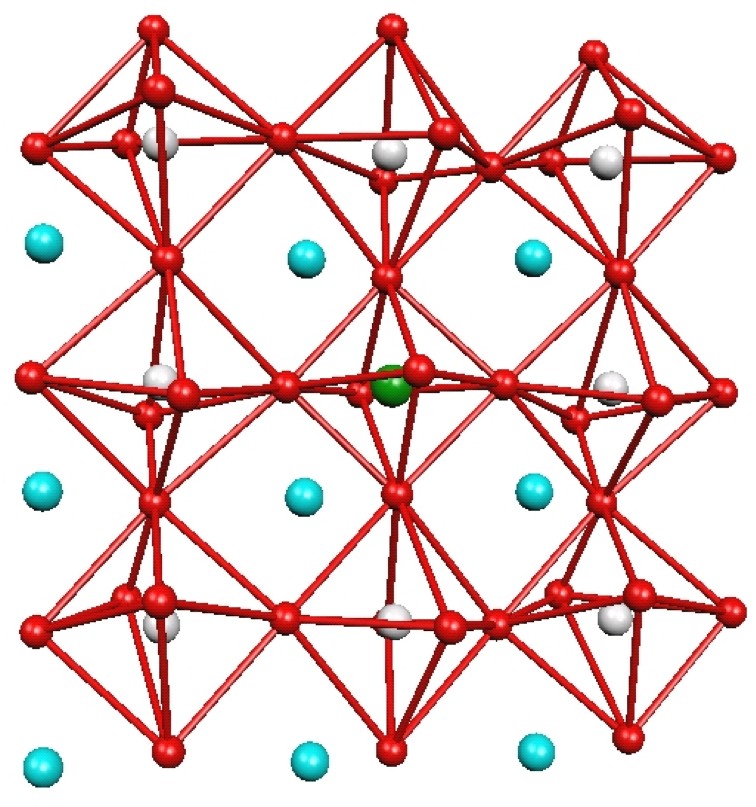}
\caption{{Structure 3, where Pd dopant is not accompanied by an O
vacancy. Pd is green, Ba are light blue, Ce are gray and O red.}}
\label{fig:Structure3}
\end{figure}

 Structures 1, 2, and 3 were modeled with $a$ = 8.799, $b$ = 8.759 and $c$ =
 8.823~\AA\ , the relaxed DFT lattice constants of
 BaCe$_{0.875}$Pd$_{0.125}$O$_{3}$ (Structure 3). Doping BaCeO$_{3}$ with
 cationic Pd reduces the symmetry of the unit cell from $Pnma$ to the space
 group $P1$. The tilt system in BaCeO$_{3}$ is severely interrupted with the
 addition of one Pd, then becomes closer to the pure BC tilt pattern with the
 addition of a Pd-O vacancy pair. In Structure 3, the substitution of one Ce
 for one Pd with no accompanying vacancy has tilt angles that range from 2 to
 15$^{\circ}$. The average tilt angle of this system is
 9.6~$\pm$~4.6$^{\circ}$. This perturbation results in a dramatic decrease in
 symmetry, also displayed by the Ce-O bond lengths. The three nearest Ce
 neighbors (Ce4, 6 and 7 in Table~\ref{table:B-Odist3}) all have two distinct
 sets of bond lengths; four bonds around 2.21~\AA\, and two long bonds around
 2.36~\AA\ . The second nearest neighbors have six similar bond lengths of
 2.28~\AA\ on average. The three nearest Ce neighbors to Pd are enclosed
 within octahedra that distort, then tilt, as shown in
 Figure~\ref{fig:Structure3}. The distorted Ce octahedra are the result of the
 formation of strong, short bonds ($\approx$ 2.03~\AA) to the small Pd$^{4+}$
 cation. The bond lengths of 2.03~\AA\ are typical of Pd$^{4+}$ in an O$_{6}$
 environment ($r_{\rm O^{2-}}$=1.42 and $r_{\rm Pd^{4+}}$=0.615).
 
\begin{figure}
\includegraphics[width=3.0in]{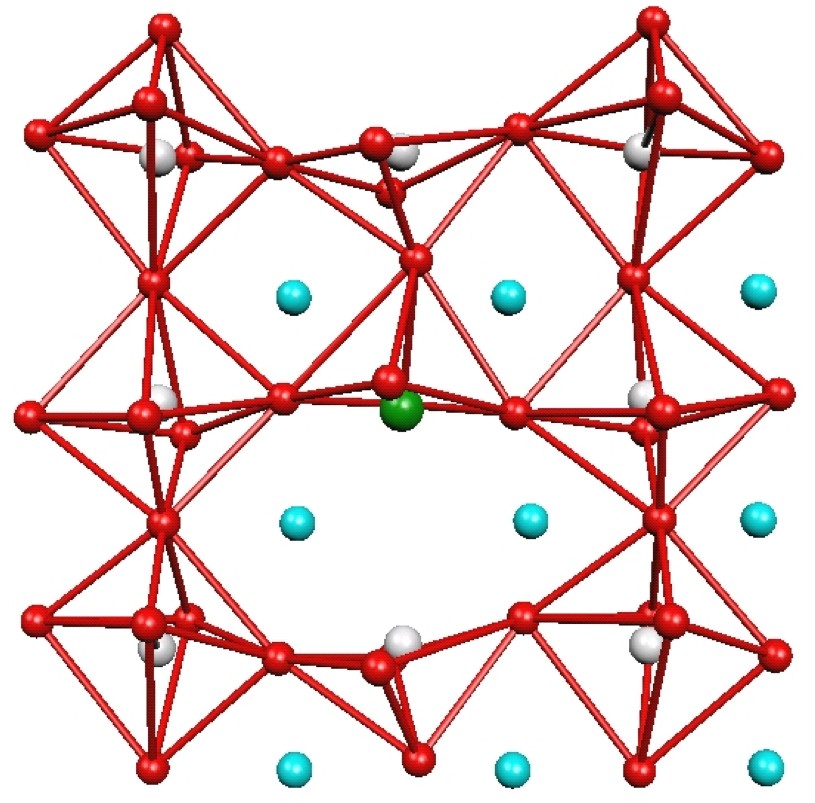}
\caption{{Structure 1, where Pd dopant is adjacent to an O
vacancy. This shows the lesser degree of octahedral distortion of the
nearest neighbors when compared to Structure 2, as well as the
increasingly symmetric tilt. Pd is green, Ba are light blue, Ce are
gray and O red.}}
\label{fig:Structure1}
\end{figure}

\begin{figure}
\includegraphics[width=3.0in]{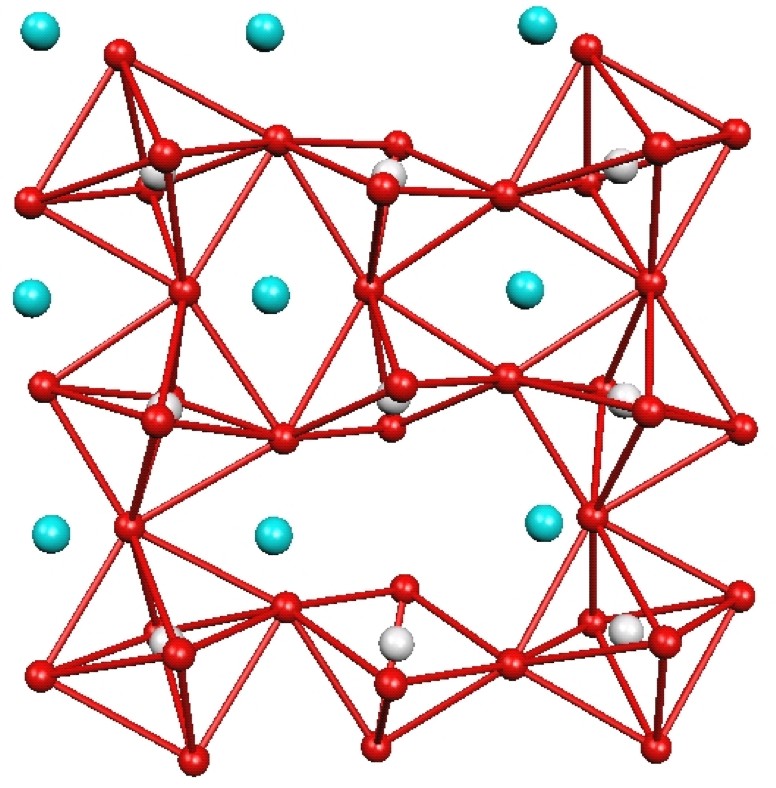}
\caption{{Structure 2 has an O vacancy between two Ce atoms, not
adjacent to Pd (lower plane, not shown). Ba are light blue, Ce are
gray and O red.}}
\label{fig:Structure2}
\end{figure}

 Of the two structures with an oxygen vacancy, the higher energy structure is
 Structure 2, where the O vacancy is in between two Ce. This is 0.95 eV higher
 in energy than Structure 1, because the reduction of Pd$^{4+}$ to Pd$^{2+}$
 is more favorable than the reduction of two Ce$^{4+}$. The vacancy in
 Structure 1 is further stabilized by the formation of a square planar
 environment around Pd$^{2+}$. As shown in Figure~\ref{fig:Structure1}, the Pd
 in Structure 1 is four-coordinate square-planar, and not five-coordinate
 square-pyramidal. In a square-planar environment, $r_{\rm
 PD^{2+}}=0.64$. This is supported by the Pd-O bond lengths in Structure 1
 from Table~\ref{table:B-Odist1}. The first four shortest bond lengths are on
 average 2.04~\AA, and the fifth is 2.66~\AA. This longer bond length occurs
 as a result of the apical O in the CeO$_{5}$ square pyramid moving away from
 the Pd. This does not occur in Structure 2, where the vacancy is in between
 two Ce (Figure~\ref{fig:Structure2}). All six Pd-O bond lengths in Structure
 2 are around 2.02 \AA, whereas the two CeO$_{5}$ surrounding the vacancy seem
 to contract inward, as seen from the bond lengths in
 Table~\ref{table:B-Odist2}. These bond lengths are similar to those seen in
 Structure 3, where there is no vacancy.

\begin{table}
\begin{tabular}{c|cccccc}
\hline
Ce1&2.26&2.27&2.27&2.27&2.27&2.32\\
Ce2&2.23&2.24&2.24&2.27&2.27&2.29\\
Ce3&2.23&2.24&2.24&2.25&2.26&2.28\\
Ce4&2.22&2.22&2.22&2.25&2.35&2.36\\
Ce5&2.24&2.24&2.25&2.25&2.27&2.28\\
Ce6&2.05&2.21&2.21&2.23&2.23&4.51\\
Ce7&2.22&2.22&2.22&2.25&2.35&2.35\\
\hline
Pd1&2.04&2.04&2.04&2.05&2.66&4.30\\
\hline
\end{tabular}
\caption{Six shortest B-O distances (in \AA), for Structure 1,
BaCe$_{0.875}$Pd$_{0.125}$O$_{2.875}$, with vacancy between Pd1 and
Ce6.}
\label{table:B-Odist1}
\end{table}

\begin{table}
\begin{tabular}{c|cccccc}
\hline
Ce1&2.21&2.22&2.22&2.22&2.24&4.34\\
Ce2&2.22&2.22&2.28&2.28&2.30&2.34\\
Ce3&2.17&2.21&2.21&2.22&2.23&4.34\\
Ce4&2.26&2.28&2.28&2.31&2.37&2.38\\
Ce5&2.22&2.22&2.26&2.27&2.32&2.35\\
Ce6&2.26&2.27&2.29&2.31&2.35&2.37\\
Ce7&2.22&2.22&2.22&2.22&2.36&2.42\\
\hline
Pd1&2.01&2.01&2.01&2.02&2.02&2.03\\
\hline
\end{tabular}
\caption{Six shortest B-O distances (in \AA), for Structure 2
  BaCe$_{0.875}$Pd$_{0.125}$O$_{2.875}$, with vacancy in between Ce1
  and Ce3.}
\label{table:B-Odist2}
\end{table}

\begin{table}
\begin{tabular}{c|cccccc}
\hline
Ce1&2.28&2.28&2.28&2.28&2.29&2.29\\
Ce2&2.23&2.24&2.26&2.26&2.27&2.27\\
Ce3&2.25&2.25&2.25&2.25&2.26&2.26\\
Ce4&2.20&2.20&2.22&2.22&2.36&2.36\\
Ce5&2.24&2.24&2.26&2.26&2.26&2.26\\
Ce6&2.21&2.21&2.21&2.21&2.37&2.37\\
Ce7&2.21&2.21&2.22&2.22&2.35&2.35\\
\hline
Pd1&2.03&2.03&2.03&2.03&2.04&2.04\\
\hline
\end{tabular}
\caption{Six shortest B-O distances (in \AA), in Structure 3
BaCe$_{0.875}$Pd$_{0.125}$O$_{3}$, with no vacancies}
\label{table:B-Odist3}
\end{table}

This clearly indicates that Pd reduction via an O vacancy may be needed to
alleviate some of the strain of this system induced by the small Pd$^{4+}$
ion. When an O vacancy is placed between two Ce, as in Structure 2, the
closest corner sharing octahedra decrease their tilt angles to either 9 or
2$^{\circ}$. The remaining octahedra slightly increase their tilt angles to
either 13 or 14$^{\circ}$. Structure 2 is more distorted than Structure 3, and
it seems to have less of a concerted tilt preference. The average tilt angle
in Structure 2 is 9.9~$\pm$~5.1$^{\circ}$. In Structure 1, when the vacancy is
between Pd and Ce, the corner sharing octahedra are not as affected,
decreasing their tilt angles by at most 2$^{\circ}$. The CeO$_{5}$ next to the
vacancy decreases its tilt angle to 4$^{\circ}$. The average tilt angle of
Structure 1 is 9.9~$\pm$~3.4$^{\circ}$. The average tilt angle of Structures 1
and 2 is the same, but the standard deviation is larger for Structure 2. The
largest interruption of concerted tilt systems by a vacancy occurs in
Structure 2, where it is not stabilized by the Pd$^{2+}$ cation. All four
supercell structures resemble the tilt system $a^{-}a^{+}a^{-}$, however
symmetry has been broken by both the Pd substitution and accompanying O
vacancy.

\begin{figure}
\includegraphics[width=3.0in]{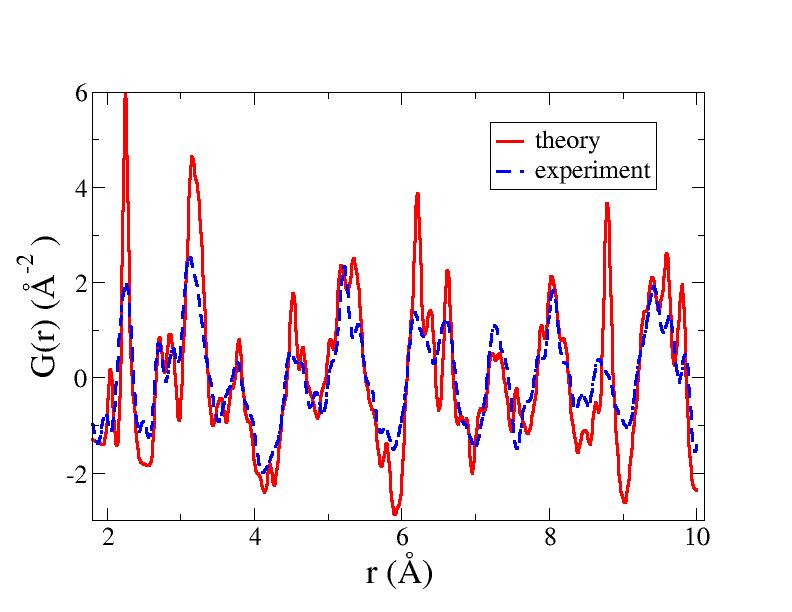}
\caption{{Experimental neutron PDF $G(r)$ of BCP, shown in dashed blue line, and
    DFT PDF $G(r)$ of Structure 1, shown in solid red line. Experimental BCP
    was of formula BaCe$_{1-x}$Pd$_{x}$O$_{3-\delta}$ where x=0.10.}}
\label{fig:PDF}
\end{figure}

To elucidate the local structure of a disordered perovskite, such as BCP, we
simulate a neutron-scattering pair distribution function (PDF) by combining
DFT-computed atomic positions and known neutron scattering factors. To make
direct comparison with experimental neutron PDF, we subtract the average
number density and multiply by $4\pi r$:

\begin{eqnarray}
 G(r) = 4\pi r (\rho(r)-\rho_{0}(r))
\end{eqnarray}            

\noindent where $\rho(r)$ is the microscopic pair density and $\rho_{0}(r)$ is
the average number density. The experimental PDF G(r) of BCP generated with
the program PDFFIT~\cite{Proffen99p572} matches closely the DFT-derived PDF of
Structure 1 generated by an in-house code used in previous
studies~\cite{Grinberg04p144118,Juhas04p214101}. The correspondence in peaks,
as shown in Figure~\ref{fig:PDF}, is strong, especially for short $r$. This
further supports the agreement between experimental data and our calculations,
as well as the assignment of Pd$^{2+}$ being stabilized by a O vacancy in
Structure 1. The match between experiment and theory was much less close for
PDFs generated from Structure 2 and Structure 3, providing evidence that
Pd-vacancy pairs are present in the experimentally prepared BCP sample.

\begin{figure}
     \centering
     \subfigure[]{
          \label{fig:4155VacOB}
          \includegraphics[width=1.5in]{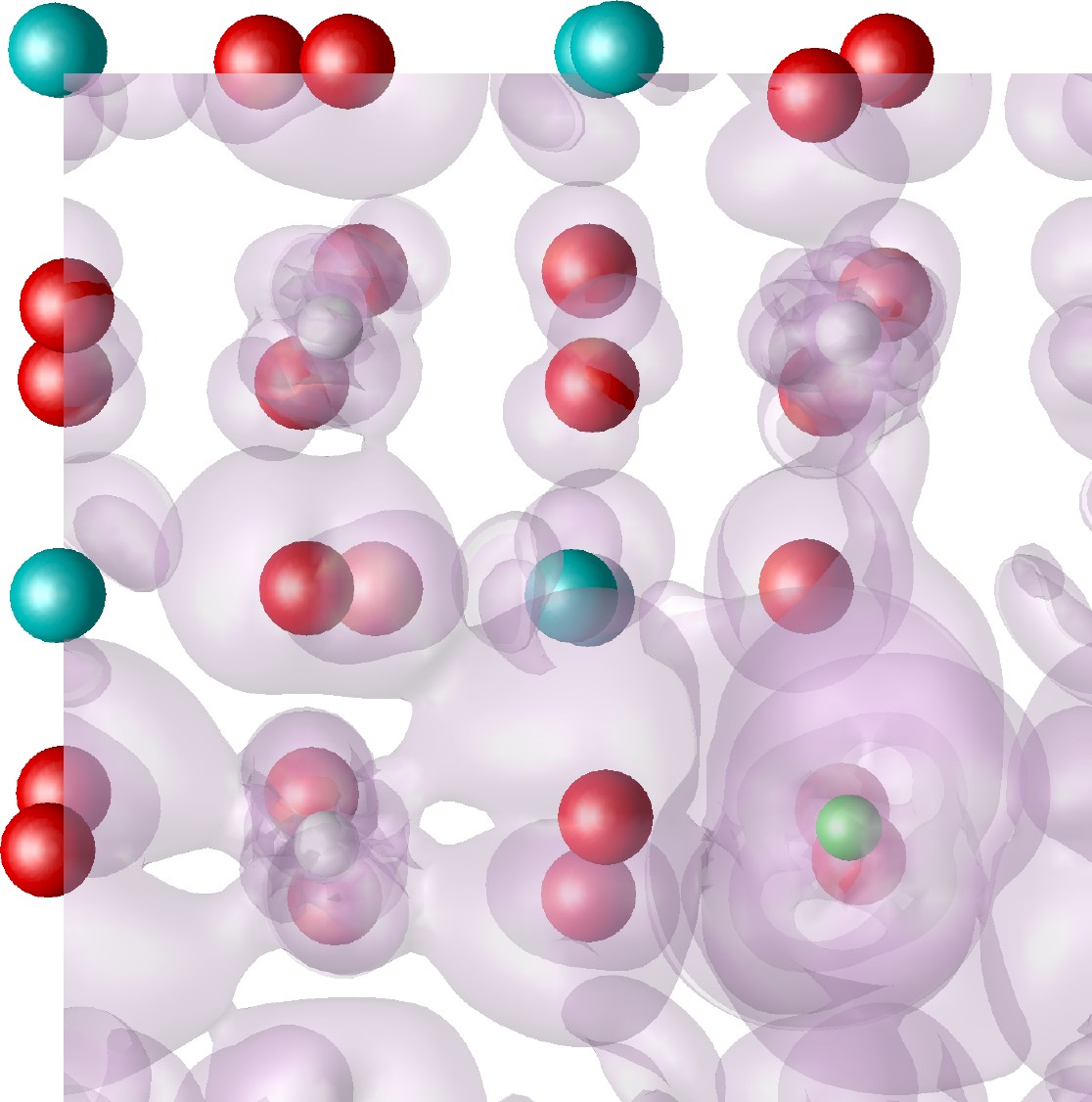}}
     \hspace{0.1in}
     \subfigure[]{
          \label{fig:4160VacOB}
          \includegraphics[width=1.5in]{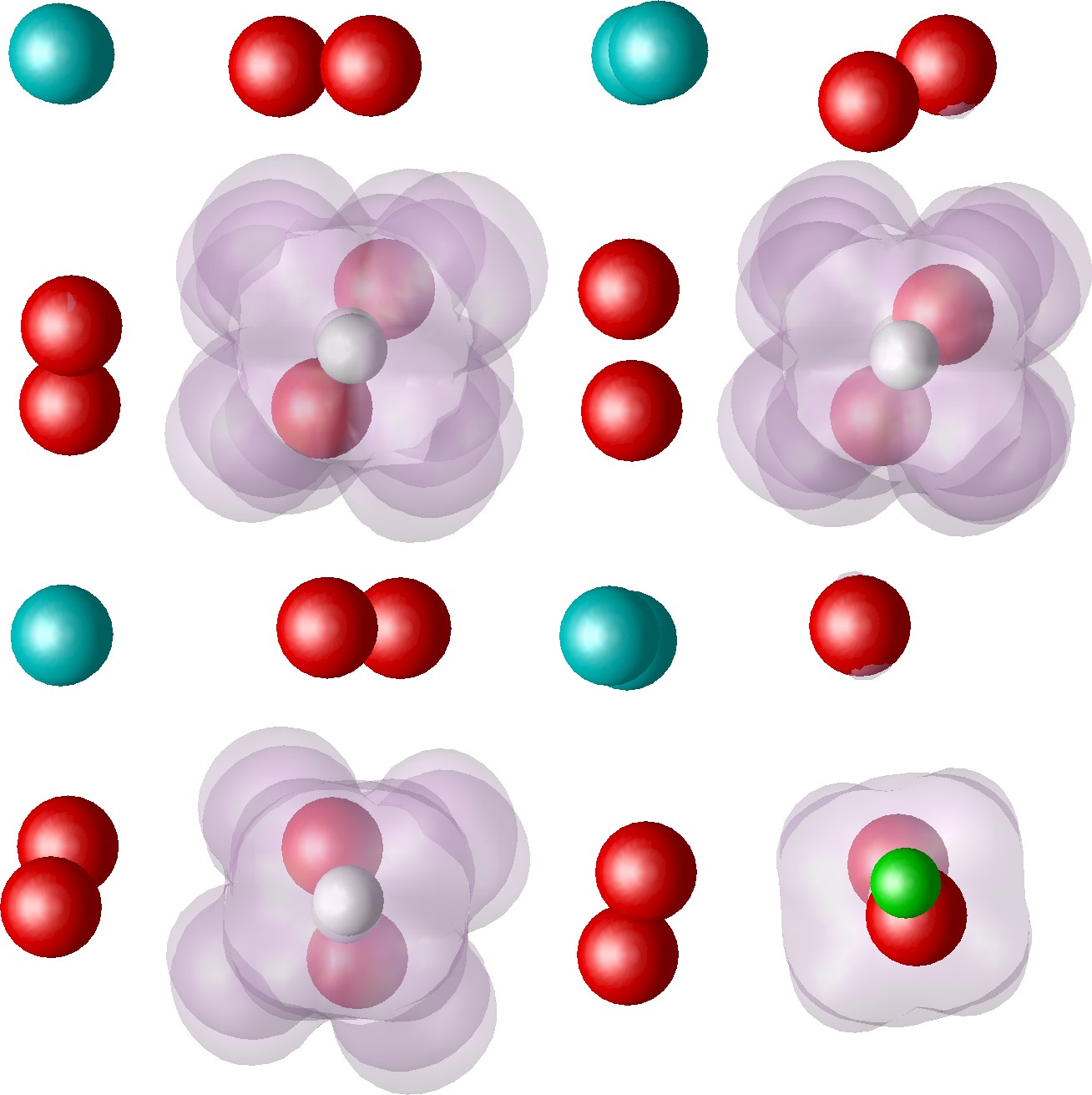}}
     \hspace{0.1in}
     \subfigure[]{
           \label{fig:4165VacOB}
          \includegraphics[width=1.5in]{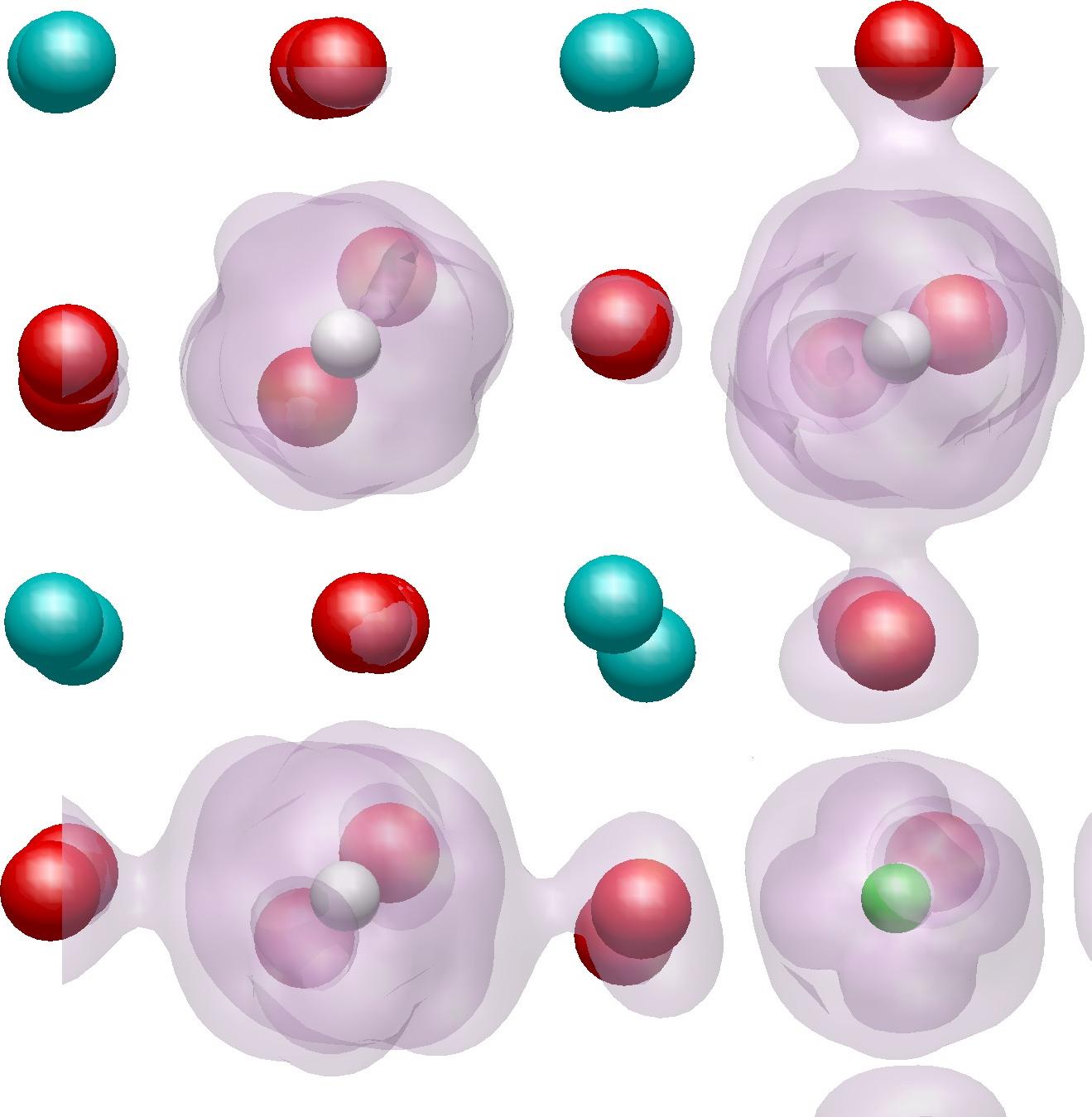}}
     \caption{{a) The HOMO of Structure 1, where the electron density
is localized around the Pd cation, and its nearest Ce neighbors. b)
The LUMO of Structure 1, composed solely of Ce $f$-orbitals. There is
no electron density localized around Pd. c) An empty state above the
LUMO, composed of Ce $f$-orbitals and a Pd $d_{x^2-y^2}$. In all
figures, Pd is the green cation located at the right corner, Ce are
white, Ba are light blue and O red. Bonds are omitted for clarity of
the isosurface shown.}}
       \label{fig:VacOBwaves}
\end{figure}

Analysis of the density of states (DOS) of the three structures shows that
Structure 1 contains an interesting feature absent from either Structures 2 or
3. Above the Fermi level, there is a $d$-orbital localized on Pd that is
$d_{x^{2}-y^{2}}$.This is shown in Figure~\ref{fig:4165VacOB} and indicates
the presence of Pd$^{2+}$. This $d$-orbital is not present in either the HOMO
(Figure~\ref{fig:4155VacOB}) or LUMO (Figure~\ref{fig:4160VacOB}). In a
square-planar environment, this is the non-degenerate $d$-orbital that is
highest in energy. It is absent from both Structures 2 and 3 because Pd instead
assumes a Pd$^{4+}$ oxidation state in octahedral coordination.

\begin{figure}
\includegraphics[width=3.0in]{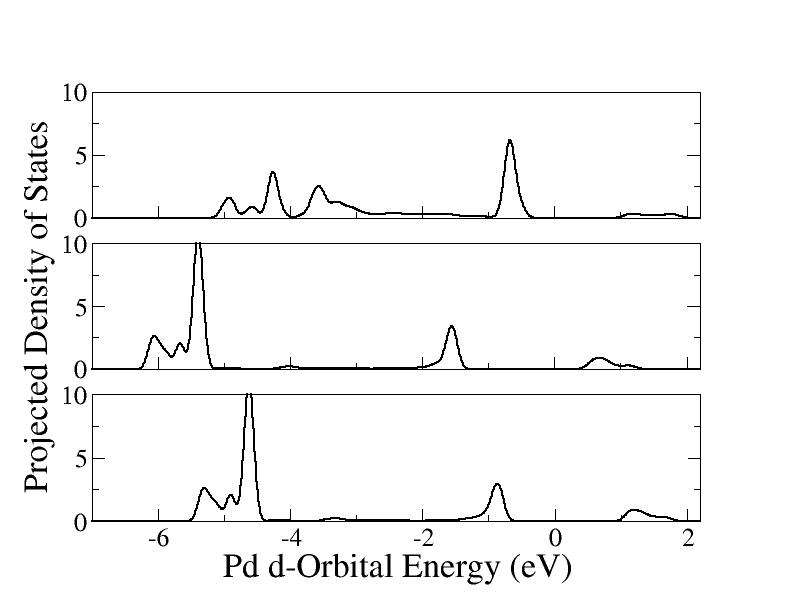}
\caption{{Projected Density of States of the Pd $d$-states. Structure 1 (top)
shows that a filled band is being stabilized when an oxygen vacancy is placed
between a Pd-Ce pair, as $d$-states above the Fermi level become less
populated. This feature is not seen in either Structures 2 (middle) or 3
(bottom). In these two structures, there appear sharply intense bands around
-5 eV, absent in Structure 1. These are filled states, lower in energy, that
demonstrate the more ionic nature of the Pd in Structures 2 and 3.}}
\label{fig:PDOScomp}
\end{figure}

A comparison of the projected density of states of the Pd $d$-orbitals in
Figure~\ref{fig:PDOScomp} shows that the Pd in Structure 1 is different than
the more similar Structures 2 and 3. Since Structures 2 and 3 both contain
Pd$^{4+}$ in an octahedral environment, they have the same $d$-states, though
shifted in energy. The Pd in Structure 2 is slightly more ionic than in
Structure 3, an effect caused by the O vacancy in Structure 2. This O vacancy,
in between two Ce, has caused the octahedron around Pd to contract slightly,
forming stronger, more ionic, Pd-O bonds. In Structure 1, new $d$-states
closer to the Fermi level are populated, and previously unfilled bands, above
the Fermi level, decrease in intensity. This is because Pd has been stabilized
in a square planar environment as Pd$^{2+}$. The shift of the $d$-states
closer towards the Fermi level when compared to Structures 2 and 3 also
support Pd being less ionic.

\section{\label{sec:level1} Conclusion}
We have presented here a DFT study of the ground state structures of the
undoped and Pd-doped rare-earth perovskite BaCeO$_{3}$. The Wyckoff positions
and space group assignment for BaCeO$_{3}$ are in agreement with
experiment. The substitution of Ce by Pd without an accompanying O vacancy
distorts the doped perovskite moreso than with an accompanying vacancy. When
the vacancy is not adjacent to Pd, but is between two Ce, the structure is
less stable. Located between a Pd-Ce pair, the vacancy causes the CeO$_{5}$
square pyramid to tilt away from Pd, stabilizing a four-coordinate
square-planar Pd geometry. The stabilization is about 0.95 eV relative to
Structure 2. The coordination of Pd is also supported in the position of
filled Pd $d$-states just below the Fermi level of Structure 1 that are not
filled in either Structures 2 or 3. This is also supported by the neutron pair
distribution function data in Reference~\cite{Li07p1418}. The stabilization of
oxygen vacancies by Pd doping leads to an enhanced rate of oxygen diffusion
through the material~\cite{Singh07p347}, which could be used in the design of
a solid oxide fuel cell membrane or low temperature catalyst.

This work has been supported by the National science Foundation (CTS05-08455,
DMR05-20415, DMR05-20020), the Department of Energy (DE-FG02-05ER15725,
DE-AC52-06NA25396, DE-xxxx-xxxxxxxxx) Office of Basic Energy Sciences (BES)
and the Air Force Office of Scientific Research
(FA9550-07-1-0397). Computational support was provided by a Challenge Grant
from the High Performance Computing Modernization Office.

\bibliography{thebibliography}

\end{document}